\documentclass[aps,twocolumn,prl,superscriptaddress,amsmath,showpacs,tightenlines,pdflatex,longbibliography]{revtex4-1}
\usepackage{xcolor}
\usepackage{amsmath}

\usepackage{graphicx,times}
\usepackage{amstext}
\usepackage{amsmath}            %serve per le subequazioni
\usepackage{amssymb}            %serve per il simbolo "marchio registrato", \circledR
\usepackage{graphicx}           %serve per le figure eps, ps etcyy
\usepackage{latexsym}
\usepackage{bm}

\usepackage{etoolbox}
\usepackage{breqn}

\usepackage[FIGTOPCAP,raggedright,nooneline]{subfigure}

\makeatletter
\let\cat@comma@active\@empty
\makeatother

\usepackage{url}
\usepackage[colorlinks]{hyperref}
\hypersetup{%
    plainpages=true,
    breaklinks=true,%not default in dvips mode, so we must specify
    hypertexnames=false,%not ideal, but needed when pagenums duplicate (`i' vs. `1')
    pageanchor=true,
    colorlinks=true,
    linkcolor={blue},
    citecolor={red},
    urlcolor={blue},
    anchorcolor={black}
}

\newcommand{\beq}{\begin{eqnarray}}
\newcommand{\eeq}{\end{eqnarray}}

\def\<{\langle}
\def\>{\rangle}

\def \info#1{}

\begin{document}

\title{Spatio-Temporal Steering for Testing Nonclassical Correlations in Quantum Networks}

\author{Shin-Liang Chen}
\affiliation{Department of Physics, National Cheng Kung
University, Tainan 701, Taiwan}
\affiliation{Naturwissenschaftlich-Technische Fakult\"at,
Universit\"at Siegen, Siegen, Germany}

\author{Neill Lambert}
\affiliation{CEMS, RIKEN, 351-0198 Wako-shi, Japan}

\author{Che-Ming Li}
\affiliation{Department of Engineering Science, National Cheng
Kung University, Tainan 701, Taiwan}

\author{Guang-Yin Chen}
\affiliation{Department of Physics, National Chung Hsing
University, Taichung 402, Taiwan} \affiliation{CEMS, RIKEN,
351-0198 Wako-shi, Japan}

\author{Yueh-Nan Chen}
\email{yuehnan@mail.ncku.edu.tw} \affiliation{Department of
Physics, National Cheng Kung University, Tainan 701, Taiwan}
\affiliation{Physics Division, National Center for Theoretical
Sciences, Hsinchu 300, Taiwan} \affiliation{CEMS, RIKEN, 351-0198
Wako-shi, Japan}

\author{Adam Miranowicz}
\affiliation{CEMS, RIKEN, 351-0198 Wako-shi, Japan}
\affiliation{Faculty of Physics, Adam Mickiewicz University,
61-614 Pozna\'n, Poland}

\author{Franco Nori}
\affiliation{CEMS, RIKEN, 351-0198 Wako-shi, Japan}
\affiliation{Department of Physics, The University of Michigan,
Ann Arbor, Michigan 48109-1040, USA}

\begin{abstract}
We introduce the concept of spatio-temporal steering (STS), which
reduces, in special cases, to Einstein-Podolsky-Rosen steering and
the recently-introduced temporal steering. We describe two
measures of this effect referred to as the STS weight and robustness.
We suggest that these STS measures  enable a new way to assess
nonclassical correlations in an open quantum network, such as
quantum transport through nano-structures or excitation transfer
in a complex biological system. As one of our examples, we apply
STS to check nonclassical correlations among sites in a
photosynthetic pigment-protein complex in the Fenna-Matthews-Olson
model.

\end{abstract}


% 03.65.Ta Foundations of quantum mechanics; measurement theory
% 03.67.Mn Entanglement measures, witnesses, and other characterizations
% 03.67.Bg Entanglement production and manipulation

\maketitle

%------------------------------------------------------------------
\emph{Introduction.---}%%%
Quantum steering is an intriguing quantum phenomenon, which enables one party
(usually referred to as Alice) to use her different
measurement settings to remotely prepare the set of quantum states
of another spatially-separated party (say Bob). This ability,
which is not achievable without quantum resources, was first
described by Schr\"{o}dinger~\cite{Schrodinger35} in his response
to the work of Einstein, Podolsky, and Rosen (EPR)~\cite{EPR35} on
quantum entanglement and the related question about the
completeness of quantum mechanics. As recently
shown~\cite{Wiseman07}, quantum steering (also refereed to as EPR
steering) is, in general, weaker than Bell's
nonlocality~\cite{Bell64,Brunner14} but stronger than quantum
entanglement~\cite{Horodecki09}. After eighty years, quantum
steering has been gradually
formulated mathematically~\cite{Reid89,Wiseman07,Pusey13,Skrzypczyk14,Piani15}
and observed experimentally ~\cite{Reid89,Cavalcanti09,Saunders10,
Walborn11,Wittmann12, Smith12, Bennet12,
Handchen12,Steinlechner13, Su13,Schneeloch13}. Other developments
include: using steering as a resource for
quantum-information processing, quantifying
steering~\cite{Skrzypczyk14,Piani15,Gallego15,Kogias15,Costa16},
clarifying its relationship to the problem of the incompatibility of
measurements~\cite{Uola14,Quintino14,Uola15,Cavalcanti16,CBLC16},
and multipartite quantum
steering~\cite{He13,Cavalcanti15,Armstrong15,Li15_2,Xiang16},
among various other generalizations and applications (see~\cite{EPR80} and references therein).

Nonclassical temporal correlations (like photon antibunching) play
a fundamental role in quantum optics research, since the
Hanbury-Brown and Twiss experiments~\cite{HBT56} and the Glauber
theory of quantum coherence~\cite{GlauberBook}. While there is as yet no clear temporal analog of
quantum entanglement, attempts at defining such have led to new ideas about quantum causality
(see, e.g.,~\cite{Leifer13,Fitzsimons15,Horsman16} and references therein). Recently,
temporal steering~\cite{Chen14} was introduced as a temporal
analog of EPR steering, which refers to a nonclassical correlation
of a single object at different times. Contrary to temporal
entanglement, temporal steering has a clear operational
meaning~\cite{Chen14,Li15,Karthik15,Mal15,Chen16,Chiu16,Karol16a,Karol16b}.
In particular, temporal steering was used for testing the security
of quantum key distribution protocols~\cite{Chen14,Karol16a} and
for quantifying the non-Markovian dynamics of open
systems~\cite{Chen16}. Recently, temporal steering was also
experimentally demonstrated~\cite{Karol16b} by measuring the
violation of the temporal inequality presented in Ref.~\cite{Chen14}.
% This inequality is a temporal analog of the Leggett-Garg
% inequality~\cite{LG85} derived to test the assumption of
%``macroscopic realism'' on a single object.
Moreover, a measure of temporal steering was
proposed~\cite{Chen16,Karol16a} and experimentally
determined~\cite{Karol16b}.

Here, we introduce the concept of spatio-temporal steering (STS)
as a natural unification of the EPR and temporal forms of steering.
In addition, we propose two measures of STS, specifically, its robustness and
weight. We also show the usefulness of STS in testing and
quantifying nonclassical correlations of quantum networks by
analyzing two examples, including the decay of nonclassical
correlations in quantum excitation transfers in the
Fenna-Matthews-Olson (FMO) protein complex, which is one of the
most widely studied photosynthetic
complexes~\cite{BlankenshipBook}. Note that STS can also be
applied to test quantum-state transfer in quantum networks like
those described in Refs.~\cite{Christandl04,Chakraborty16}.

%------------------------------------------------------------------
\emph{Temporal steering: From temporal hidden-variable model to temporal hidden-state model.---}%%%
Let us briefly review the so-called temporal hidden-state model
for a single system at two moments of
time~\cite{Chen14,Li15,Chen16}. Consider that, during the
evolution of the system from time $0$ to time $t$, one can perform
measurements using different settings $\{x\}$ and $\{y\}$ to
obtain outcomes $\{a\}$ and $\{b\}$ at times 0 and $t$,
respectively. If one makes two assumptions: (A1)
noninvasive measurability at time $0$, which means that one can
obtain a measurement outcome without disturbing the system, and
(A2) macrorealism (macroscopic realism)~\cite{LG85}, which means
that the outcome of the system pre-exists, no matter if a measurement
has been performed or not. Under these conditions, there exist
some hidden variables $\lambda$, which \emph{a priori} determine the joint
probability distributions $p(a,b|x,y) = \sum_\lambda
p(\lambda)p(a|x,\lambda)p(b|y,\lambda)$~\cite{Fritz10,Dressel11,Maroney12,Kofler13,Budroni13,Emary14}.

Now, if one replaces the assumption (A2) with (A2'), which means
that during each moment of time the system can be described by a
quantum state $\sigma_\lambda$, which is determined by some hidden
variables $\lambda$ independent of the measurements performed
before, then the hidden variables determine not only the observed
data table $p(a|x)=\sum_\lambda p(\lambda)p(a|x,\lambda)$ at time
$t=0$, but also \emph{a priori} the quantum state $\rho =
\sum_\lambda p(\lambda)\sigma_\lambda$ at time $t$. It is convenient
to define the \emph{temporal assemblage}
$\{\sigma_{a|x}^{\text{T}}(t)\equiv
p(a|x)\tilde{\sigma}_{a|x}(t)\}_{a,x}$, where
$\tilde{\sigma}_{a|x}(t)$ is the observed quantum state at time
$t$ conditioned on the earlier measurement event $a|x$ at time
$0$. Thus, the temporal assemblage is a set of subnormalized
states, which characterizes the \emph{joint behaviour}: (1)
$p(a|x)=\text{tr}[\sigma_{a|x}^{\text{T}}(t)]$ and (2)
$\tilde{\sigma}_{a|x}(t) = \sigma_{a|x}^{\text{T}}(t) /
\text{tr}[\sigma_{a|x}^{\text{T}}(t)]$. Furthermore, the
formulation of the temporal hidden-state model can be written as
$\sigma_{a|x}^{\text{T}}(t) = \sum_\lambda
p(\lambda)p(a|x,\lambda)\sigma_\lambda$. Quantum mechanics
predicts some assemblages, which do not admit the temporal
hidden-state model, and we refer to this situation as
\emph{temporal steering}~\cite{Chen16}. Note that since the
hidden-state model is a strict subset of the hidden-variable
model, using the former model may admit an easier detection of the
nonclassicality of the quantum dynamics than using the
hidden-variable
 model.

\begin{figure}
\begin{center}
% \subfigure[ ]
{\includegraphics[width=0.4\textwidth]{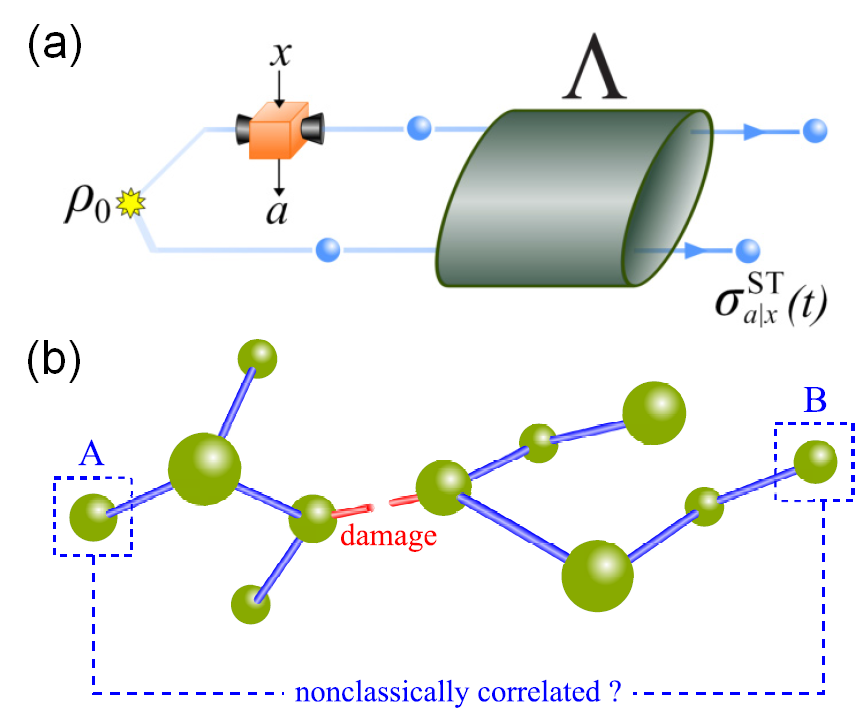}}\\%\hfill
\caption{(a) Schematic diagram of  spatio-temporal steering. At
time $t=0$, a system A (which may be entangled with a system B) is
subject to a local measurement with one of the measurement
settings $\{x\}$, which is described by a positive-operator-valued
measure $\{F_{a|x}\}_a$. After this measurement, the
post-measurement composite state $\protect\rho _{a|x}$ is sent
into a quantum channel $\Lambda$ and evolves for a time period
$t$. After many rounds of the experiment, the set of subnormalized
quantum states of the system B is denoted as
$\{\sigma_{a|x}^{\text{ST}}(t)\}_{a,x}$. With some appropriate
$\rho_{0}$, $\{F_{a|x}\}_{a,x}$, and $\Lambda$, the assemblage
$\{\sigma_{a|x}^{\text{ST}}(t)\}_{a,x}$ does not admit the
spatio-temporal hidden-state model Eq.~(\ref{HSmodel}). We call
this \emph{spatio-temporal steering} and refer the assemblage
$\{\sigma_{a|x}^{\text{ST}}(t)\}_{a,x}$ as \emph{spatio-temporal
steerable}. (b)  A schematic example of a quantum network with
damage (strong dissipation or dephasing, or an entirely broken
link).  The STS weight and robustness can be employed as
diagnostic tools to check whether site-A and site-B are
nonclassically correlated. } \label{sch_dia}
\end{center}
\end{figure}

\emph{Spatio-temporal steering.---}%%%
Similarly, we can also generalize the hidden-state model to the
hybrid spatio and temporal scenario. That is, we would like to
consider the hidden-state model for a system B at time $t$, after
the local measurement has been performed on a system A at time
$0$. Then, under the assumptions of non-invasive measurement for
the system A at time $0$ and the hidden state for the system B at
time $t$, the spatio-temporal hidden-state model is written as
(for brevity, the term ``spatio-temporal'' will be sometimes
omitted hereafter.)
\begin{equation}%%
\sigma_{a|x}^{\text{ST,B}}(t) = \sum_\lambda
p(\lambda)p_{\text{A}}(a|x,\lambda)\sigma_\lambda^{\text{B}} ~~~~~~~~\forall ~a,x,
\label{HSmodel}
\end{equation}
where $\sigma_{a|x}^{\text{ST,B}}(t)\equiv p_{\text{A}}(a|x)
\tilde{\sigma}_{a|x}^{\text{ST,B}}(t)$, with
$\tilde{\sigma}_{a|x}^{\text{ST,B}}(t)$ being the observed quantum
state of the system B at time $t$, conditioned on the measurement
event $a|x$ [with corresponding data table $p_{\text{A}}(a|x)$] of
the system A at time $0$. When there is no risk of confusion, we
will abbreviate $\sigma_{a|x}^{\text{ST,B}}(t)$ as
$\sigma_{a|x}^{\text{ST}}(t)$, $p_{\text{A}}(a|x)$ as $p(a|x)$,
and $\sigma_\lambda^{\text{B}}$ as $\sigma_\lambda$. The set of
subnormalized states $\{\sigma_{a|x}^{\text{ST}}(t)\}_{a,x}$ is
refereed to as a \emph{spatio-temporal assemblage} having the
property $p(a|x)=\text{tr}[\sigma_{a|x}^{\text{ST}}(t)]$ and
$\tilde{\sigma}_{a|x}^{\text{ST}}(t)=\sigma_{a|x}^{\text{ST}}(t) /
\text{tr}[\sigma_{a|x}^{\text{ST}}(t)]$, and can be certified if
it admits the model, given by Eq.~(\ref{HSmodel}), via the
following semidefinite programming (SDP) (see \cite{SDP} for SDP,
and~\cite{Pusey13,Skrzypczyk14,Cavalcanti16} for dealing with the
certification of the hidden-state model for a given assemblage):
\begin{equation}%%
\begin{aligned}
\text{find}~~& \{\rho_\lambda\} \\
\text{subject to}~~&\sigma_{a|x}^{\text{ST}}(t) = \sum_{\lambda}p(a|x,\lambda)\rho_\lambda &&\forall ~a,x,\\
&\text{tr}\sum_\lambda\rho_\lambda = 1, ~~~~~~~~\rho_\lambda\geq 0
&&\forall ~\lambda,
\end{aligned}
\label{SDP_HSmodel}
\end{equation}
where $\rho_\lambda\equiv p(\lambda)\sigma_\lambda$, and the
notation $\rho_\lambda\geq 0$ denotes that $\rho_\lambda$ is a
positive-semidefinite operator. Quantum mechanics predicts that
$\sigma_{a|x}^{\text{ST}}(t)\stackrel{\mathcal{Q}}{=}\text{tr}_{\text{A}}\left\{
\Lambda \left[(\sqrt{F_{a|x}}\otimes\openone)\rho
_{0}(\sqrt{F_{a|x}}\otimes\openone)\right] \right\}$, with
$\rho_0$ being the initial quantum state shared by the systems A
and B at time $0$, $\{F_{a|x}\}_a$ being the
positive-operator-valued measure representing the measurement $x$.
The quantum channel $\Lambda$ describes the time evolution of the
post-measurement composite system from time $0$ to time $t$ [see
the schematic diagram in Fig.~\ref{sch_dia}(a)].

%------------------------------------------------------------------

%------------------------------------------------------------------
With an appropriately designed  $\rho_0$, $\{F_{a|x}\}_{a,x}$, and
$\Lambda$, the assemblage cannot be written in the form of
Eq.~(\ref{HSmodel}) [i.e., there is no feasible solution of the
SDP problem given in Eq.~(\ref{SDP_HSmodel})]. In this situation,
the assemblage is said to be \emph{spatio-temporal steerable}. To
quantify the degree of such steerability, we would like to
introduce the quantifier called the \emph{STS weight}
($\mathcal{STSW}$), which is defined as $\mathcal{STSW} =
\text{min}~(1-\mu)~\text{subject to}~\{\sigma_{a|x}^{\text{ST}}(t)
=
\mu\sigma_{a|x}^{\text{ST,US}}(t)+(1-\mu)\sigma_{a|x}^{\text{ST,S}}(t)\}_{a,x}$
(the same techniques have been demonstrated in
Refs.~\cite{Skrzypczyk14,Chen16}).
$\{\sigma_{a|x}^{\text{ST,US}}(t)\}_{a,x}$ stands for the
unsteerable (US) assemblage [i.e., one admits
Eq.~(\ref{HSmodel})], $\{\sigma_{a|x}^{\text{ST,US}}(t)\}_{a,x}$
represents the steerable assemblage, and $0\leq\mu\leq 1$. This
can be formulated as the following SDP problem:
\begin{equation}%%
\begin{aligned} \mathcal{STSW} = \text{min}~~ &\left(1-\text{tr}\sum_{\lambda}\rho_{\lambda}\right),\quad {\rm with}\quad\rho_{\lambda}\geq 0 &&\forall ~\lambda \\
\text{subject to }~~ &\sigma_{a|x}^{\text{ST}}(t) -
\sum_{\lambda}p(a|x,\lambda)\rho_{\lambda}\geq 0 ~~&&\forall
~a,x. \end{aligned}
\label{SDP_STSW}
\end{equation}
In addition, we would like to introduce another measure, referred to as the
\emph{STS robustness} ($\mathcal{STSR}$), which can be viewed as a
generalization of the EPR steering robustness~\cite{Piani15} to
the present spatio-temporal scenario. The STS robustness
$\mathcal{STSR}$ can be defined as the minimum noise
$\tau_{a|x}^{\text{ST}}(t)$ to be added to
$\sigma_{a|x}^{\text{ST}}(t)$, such that the mixed assemblage is
unsteerable. That is, $\mathcal{STSR} =
\text{min}~\alpha~\text{subject
to}~\{\frac{1}{1+\alpha}\sigma_{a|x}^{\text{ST}}(t) +
\frac{\alpha}{1+\alpha}\tau_{a|x}^{\text{ST}}(t)=\sigma_{a|x}^{\text{ST,US}}\}_{a,x}$.
This can also be formulated as an SDP problem. Specifically,
\begin{equation}%%
\begin{aligned} \mathcal{STSR} = \text{min}~~ &\left(\text{tr}\sum_{\lambda}\rho_{\lambda}-1\right),\quad {\rm with}\quad\rho_{\lambda}\geq 0 &&\forall ~\lambda \\
\text{subject to}~~ &\sum_{\lambda}p(a|x,\lambda)\rho_{\lambda} - \sigma_{a|x}^{\text{ST}}(t) \geq 0 ~~&&\forall~a,x.
\end{aligned}
\label{SDP_STSR}
\end{equation}
The STS robustness and weight, analogously to their EPR
counterparts, have different operational meanings and properties.
For example, one could expect that these measures can imply
different orderings of states, analogously to this property
exhibited by various measures of
entanglement~\cite{Eisert99,Virmani00,Adam04}, Bell
nonlocality~\cite{Karol13}, and nonclassicality~\cite{Adam15}. A
detailed comparison of these two STS measures will be given
elsewhere~\cite{Ku16}. Here, we have calculated the STS weight
for Example~1, and the STS robustness for Example~2 in the
following sections, just to show that these measures can easily be
computed and interpreted.

\emph{Examining nonclassical correlations within a quantum network.---}%%%
A possible application of STS is that it can be used to witness
whether two nodes of a quantum network are \emph{nonclassically
correlated} (or \emph{quantum connected}). Consider two qubits on
the opposite ends of a quantum network, as shown in Fig.~1(b).
There may be a damage somewhere in the network, such that the
quantum coherent interaction between distant nodes may be
inhibited. To verify this, one can initially perform measurements
at time $t_{\text{A}}=0$ on site-A. On site-B, one performs
measurements at a later time $t$. If the value of the STS weight
(or, equivalently, the STS robustness) is always zero for the
whole range of time $t$, one can say that the influence of the
quantum measurement at site-A is not transmitted to site-B in a
steerable way.
%------------------------------------------------------------------
\begin{figure}
\begin{center}
% \subfigure[ ]
{\hspace{5mm}\includegraphics[width=0.395\textwidth]{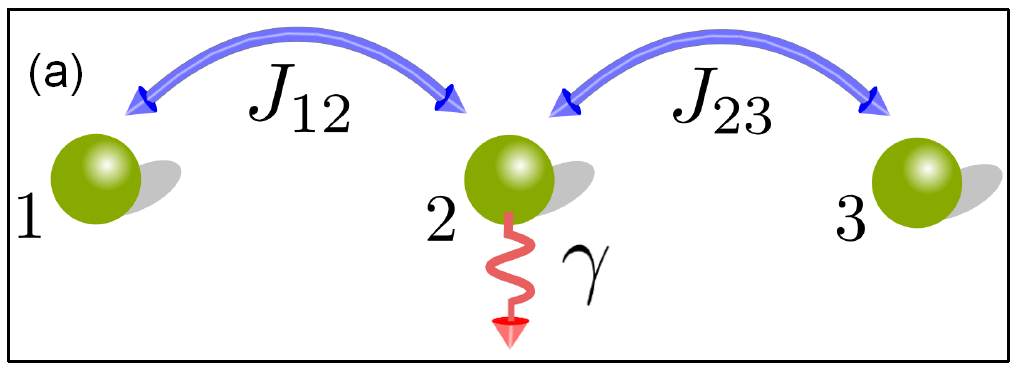}}\\%\hfill
% \subfigure[ ]
{\includegraphics[width=0.5\textwidth]{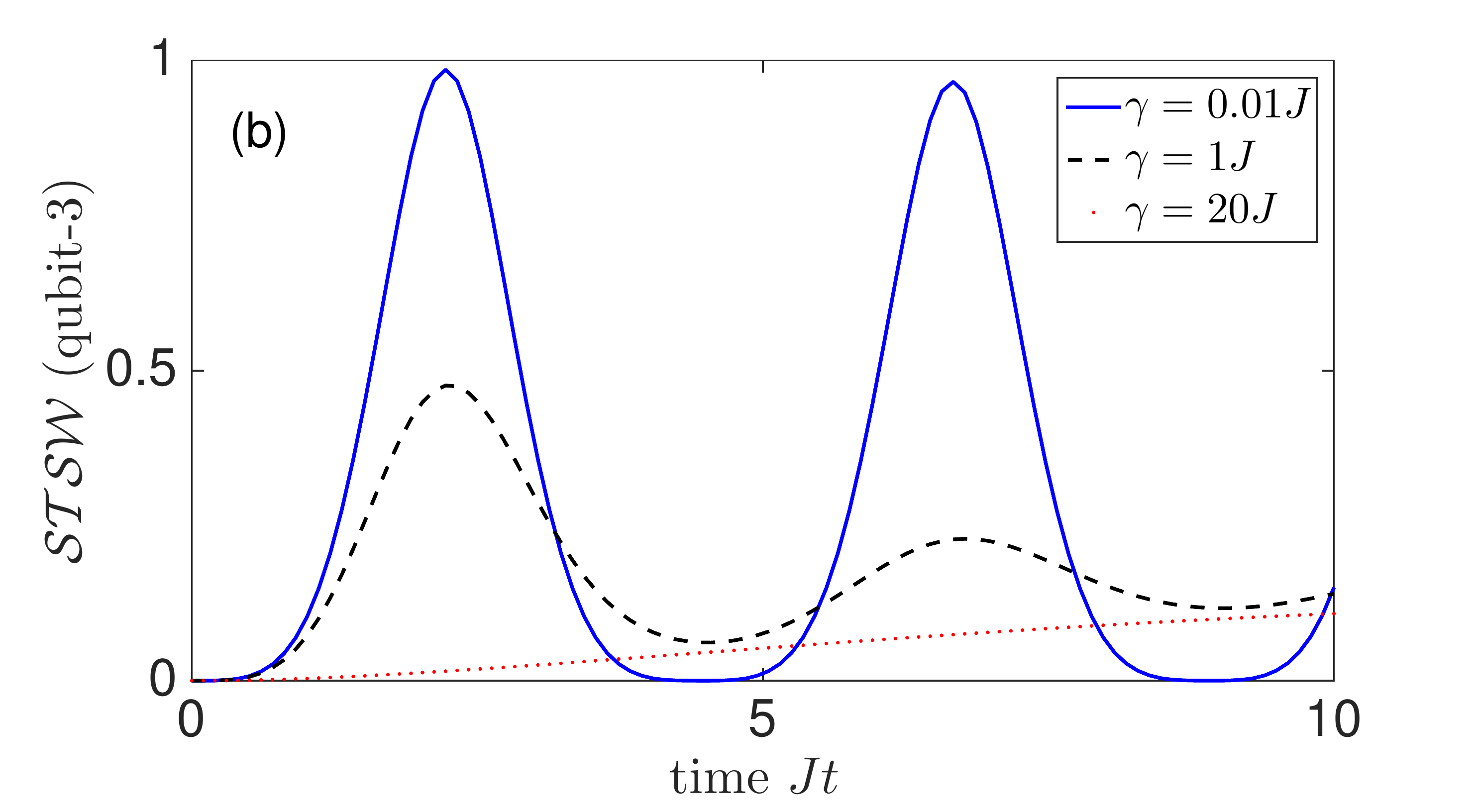}} \caption{The
STS weight versus time in a simple quantum network model
[Fig.~2(b), example~1 in the text]. (a) Three identical qubits,
with coherent coupling $J_{12}$ ($J_{23}$) between qubit 1 (2) and
2 (3). To simulate the damaged node, we assume qubit 2 suffers a
phase damping $\gamma$. (b) The blue-solid, black-dashed, and
red-dotted curves show the STS weight ($\mathcal{STSW}$) of the
assemblage $\{\sigma_{a|x}^{\text{ST}}(t)\}_{a,x}$ of qubit-3 for
different dephasing rates of the middle qubit $\gamma/J =0.01$,
$1$, and $20$, respectively. The measurement settings $\{x\}$ on
qubit-1 at time $0$ are the Pauli set ${X}$,  ${Y}$, and ${Z}$.
The initial condition is $|1\rangle \otimes |0\rangle \otimes
|0\rangle $, and $J_{12}=J_{23}\equiv J$. The time $t$ is in units
of $J^{-1}$. For brevity, we are omitting analogous plots for the
STS robustness.}
\end{center}
\end{figure}

%------------------------------------------------------------------
\emph{Example 1: The spatio-temporal steering weight in a
three-qubit network.---} As an example of STS in a quantum
network, let us apply a simplified model of two qubits coherently
coupled via a third qubit [Fig.~2(b)]. The interaction Hamiltonian of
the entire system is
\begin{equation}%%
H_{\text{int}}\ =\hbar J_{12}(\sigma _{+}^{1}\sigma
_{-}^{2}+\sigma _{-}^{1}\sigma _{+}^{2})+\hbar J_{23}(\sigma
_{+}^{2}\sigma _{-}^{3}+\sigma _{-}^{2}\sigma _{+}^{3}),
\end{equation}
where $\sigma _{+}^{i}$ ($\sigma _{-}^{i}$) is the raising
(lowering) operator of the $i$th qubit respectively, while
$J_{12}$ ($J_{23}$) is the coupling strength between qubits 1 (2)
and 2 (3). To simulate the damage in the network, and quantify it,
we assume qubit 2 may suffer noise-induced dephasing. For
simplicity, the two coupling strengths are equal, i.e.,
$J_{12}=J_{23}\equiv J$. The STS weight, calculated as described
above, is plotted in Fig.~2(b). We can see that if the dephasing
rate $\gamma $ is very small, the STS weight oscillates with time
$t$, revealing the coherent interaction between qubits 1 and 3 via
the middle qubit. If $\gamma $ is large (i.e. the middle node is
damaged), one sees the growth of the STS weight at a later time.
One can imagine that if the dephasing is very strong, it can
inhibit the appearance of the STS weight. However, several caveats arise
 in that the apparent correlations may be transmitted via
other means than the network itself (via some environment or
eavesdropper).   A possible opening for future research in this
area is to consider a multi-partite extension, and whether it can
be used as a measure of quantum communities in
networks~\cite{comm}.

\begin{figure}
\includegraphics[width=1.0\columnwidth]{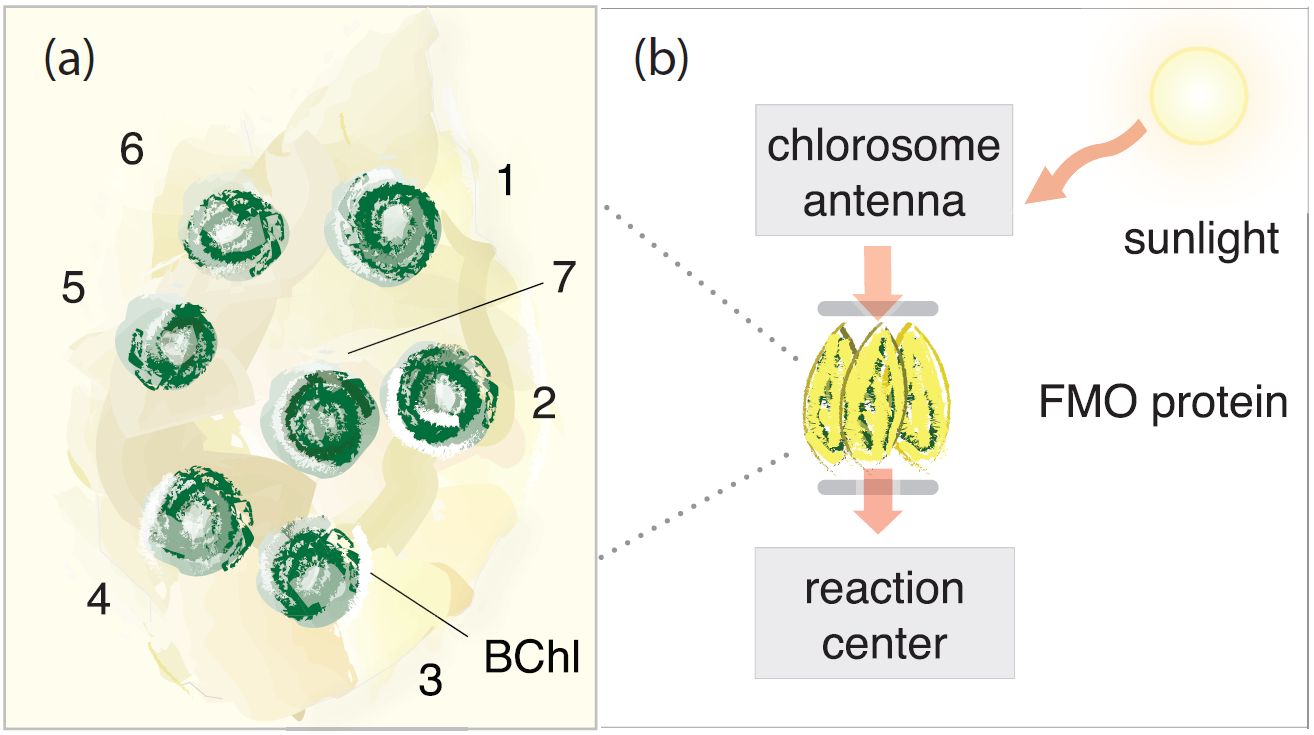}

\caption{(a) Schematic diagram of a single monomer
of the FMO protein complex. This monomer contains eight sites
(here we show only seven of them). In the bacterial
photosynthesis, the excitation from the light-harvesting antenna
enters the FMO complex at sites 6 or 1 and is then transferred
from one site to another. The excitation can irreversibly jump to
the reaction center, when it reaches site-3. In this work, the
initial condition is set as site-6 in a mixed excited state while
the other sites are in ground states. BChl stands for a
bacteriochlorophyll molecule. (b) Schematics of how the monomer
exists in a trimer, and acts as a wire connecting a large antenna
complex to the reaction center.}
\label{fig3}
\end{figure}

%------------------------------------------------------------------
\emph{Example 2: The spatio-temporal steering robustness in the
Fenna-Matthews-Olson complex.---}%%%
Much attention has been devoted to the possible functional role
of quantum coherence~\cite{Lambert13,Scholes11} in photosynthesis
bacteria, since the observation of possible quantum coherent motion of an
excitation within the FMO complex -- a photosynthetic
pigment-protein
complex~\cite{Engel07,Collini10,Panitchayangkoon10}. A simple
treatment of the excitation transfer in the FMO complex normally
considers seven coupled sites (chromophores), as shown in
Fig.~\ref{fig3}, and their interaction with the environment. The hierarchy
method~\cite{Ishizaki09a,Ishizaki09b,Ishizaki05,Tanimura90,Tanimura89}
or other open-quantum system models~\cite{Jang08,Kolli11} can be
used to explain the presence of quantum coherence and predict the
physical quantities observed in experiments.

%------------------------------------------------------------------
%------------------------------------------------------------------
\begin{figure}
% \subfigure[ ]
{\includegraphics[width=0.45\textwidth]{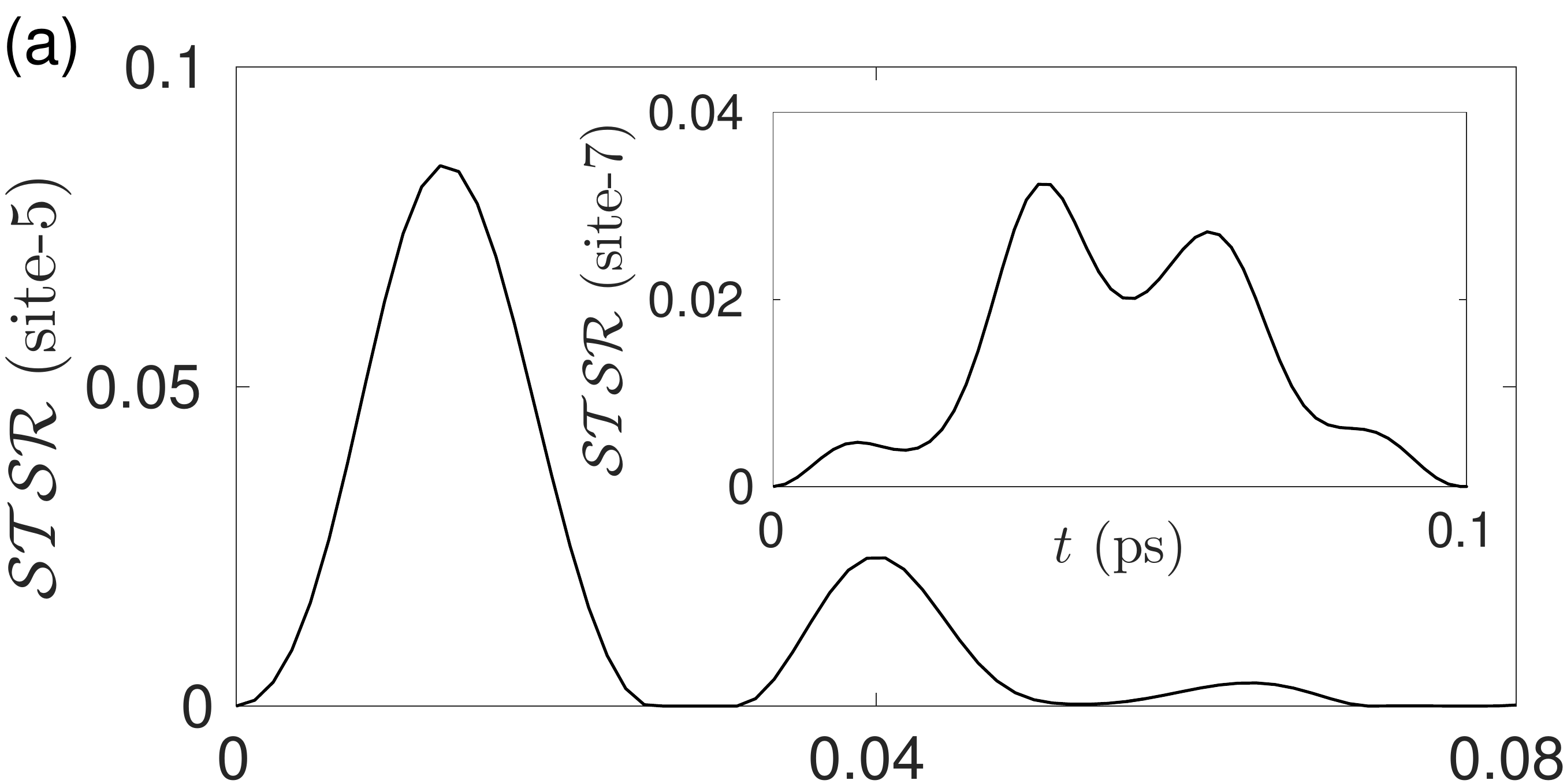}}\\%\hfill
% \subfigure[ ]

\hspace*{3mm}{\includegraphics[width=0.48\textwidth]{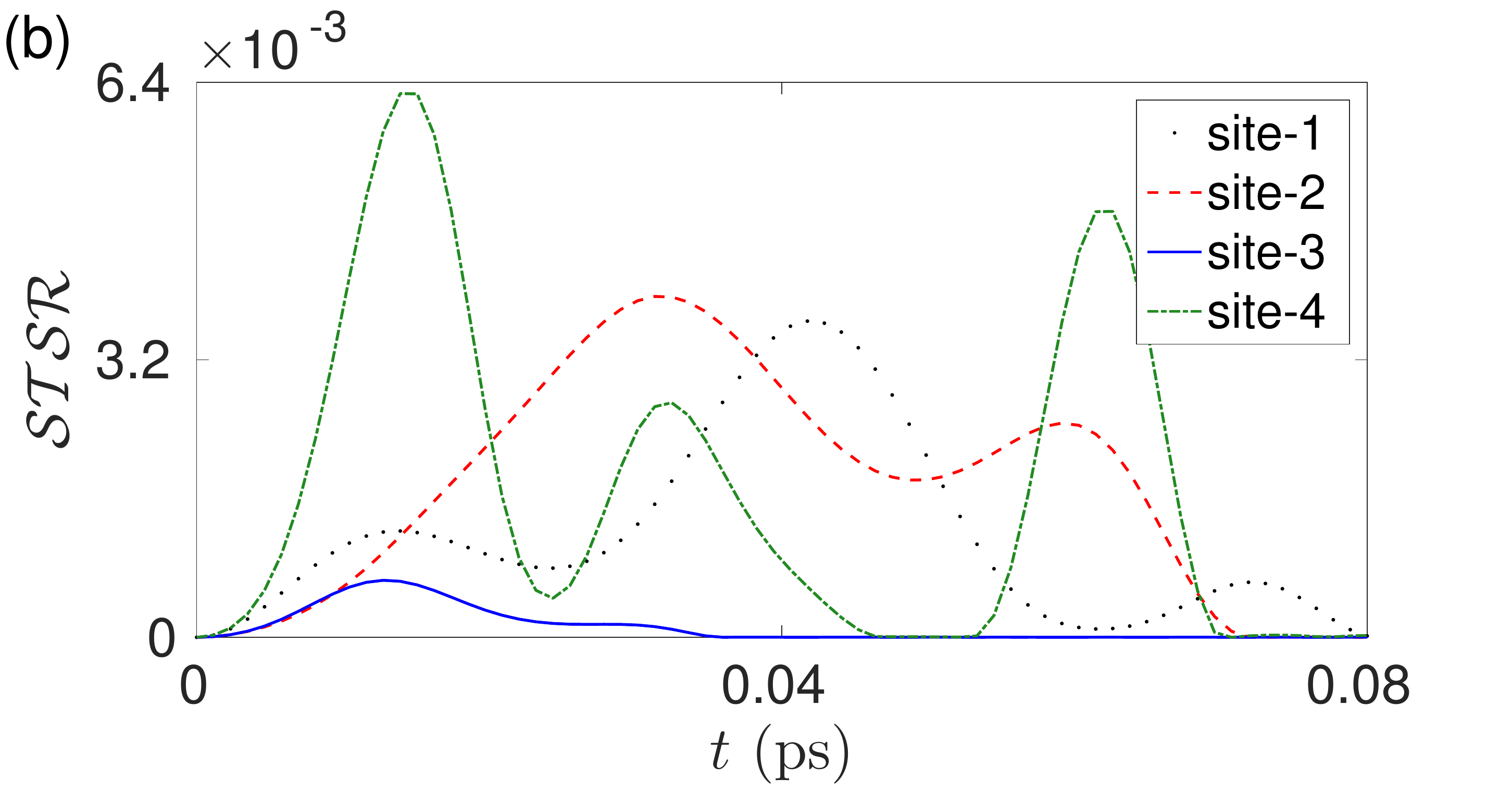}}
\caption{Evolution of the STS robustness ($\mathcal{STSR}$) in the
FMO complex. (a) The main figure together with the insets show the
decays of the STS robustness of the assemblages
$\{\sigma_{a|x}^{\text{ST}}(t)\}_{a,x}$ of site-5 and site-7
respectively. (b) The black-dotted, red-dashed, blue-solid, and
green dash-dotted curves are represent STS robustness of the
assemblages $\{\sigma_{a|x}^{\text{ST}}(t)\}_{a,x}$ of site-1, 2,
3, and 4 respectively. As the previous case, the measurement
settings on site-6 at time $0$ are the Pauli set ${X}$, ${Y}$, and
${Z}$. We assumed that the FMO is cooled down to $T=15$~K, the FMO
initial state is completely mixed at site-6 while the other sites
are in ground state, the dephasing rate is 7.7/8~cm$^{-1}$,
and the decay rate is 5.3~cm$^{-1}$. Again, for brevity, we do not
present analogous plots for the STS weight. } \label{fig4}
\end{figure}

%------------------------------------------------------------------

Empowered by STS, one can ask the following questions for a
network like the FMO protein complex: When an excitation arrives
at site-6, and propagates through the network, how large is its
quantum influence, if any, to other sites? When do such
nonclassical correlations vanish? Previously, quantum entanglement
in the FMO complex has been theoretically
analyzed~\cite{FMOentanglement}. Given the fact that the
excitation transfer is \textit{dynamic} in nature, with a specific
starting site (site-1 or site-6), it is more natural to examine
the nonclassical correlation between sites at different times by
using the STS measures. However, we point out that evaluating
these measures requires measurements in different ``excitation''
bases at both source and target sites.  Thus,  evaluating these
measures represents an analysis of the network itself, and how
quantum correlations propagate through it, much akin to the
approach taken in~\cite{biamonte14}.

In Fig.~\ref{fig4}, we numerically calculated the STS robustness
of site-6 to other sites by using the Haken-Strobl equation of
motion~\cite{MohseniBook,Chen13} (see the Supplementary
Material~\cite{Supplement}). In plotting this figure, the
temperature is chosen to be $T=15$~K with the corresponding
dephasing rate $\gamma_{\text{dp}}=7.7/8$~cm$^{-1}$ and the
decay rate (into the reaction center from site-3 only) $\Gamma=5.3
$~cm$^{-1}$.  As seen from this figure, the largest STS robustness
occurs from site-6 to site-5. This is because site-6 and site-5
have the second largest intersite coupling
($\approx89.7$~cm$^{-1}$) in the whole network. Another
interesting fact is that the robustness of site-6 to site-7 has
the second largest magnitude (with a time delay) and the longest
\textit{vanishing time} (death time) of the STS robustness. In
view of the coupling strength of the Hamiltonian, this may be due
to the relative strong couplings of site-5 to site-4
($\approx$70.7~cm$^{-1}$) and site-4 to site-7
($\approx$61.5~cm$^{-1}$), such that the influence from site-6 is
transferred through these sites with a time delay. In other words,
the STS robustness not only gives the magnitude of the
nonclassical correlations between two sites, but also gives the
information of how long the nonclassical correlation takes to
arrive, and how long it can be sustained.

\emph{Conclusions.---}%%%
Although the concept of spatio-temporal quantum entanglement is
fundamentally difficult to be described consistently, we showed
that STS, describing a certain type of spatio-temporal
nonclassical correlations, can indeed be defined and quantified in
an operational way. We hope that this may provide a wider view
than the purely spatial or temporal correlations separately. In
addition, we showed that STS, with its measures, including the STS
weight and STS robustness, can be useful to assess nonclassical
correlations in quantum networks or other open quantum systems. As
an application, we described two examples of testing nonclassical
correlations in a toy model of a three-qubit quantum network and
in a more realistic model of the excitation transfer in the
seven-site FMO complex. We believe that STS can be useful also for
testing nonclassical correlations of more complex biological
systems~\cite{Lambert13, biamonte14} and for describing quantum
transport through artificial nano-structures~\cite{neill}.
Finally, we mention that a possible experimental demonstration of
STS can be based on a delayed-time modified version of the
experiment on temporal steering reported in Ref.~\cite{Karol16b}.

%------------------------------------------------------------------
\begin{acknowledgments}%% adding this section can avoid the system counting the words in Acknowledgements.
\emph{Acknowledgements.---}%%%
The authors acknowledge fruitful discussions with Huan-Yu Ku and
Karol Bartkiewicz. We acknowledge the support of a grant from the
John Templeton Foundation. This work is supported partially by the
National Center for Theoretical Sciences and Ministry of Science
and Technology (MOST), Taiwan, grant number MOST
103-2112-M-006-017-MY4. S.-L.C. acknowledges the support of the
DAAD/MOST Sandwich Program 2016 No. 57261473. C.-M.L. and G.-Y.C.
are supported by the Ministry of Science and Technology, Taiwan,
under the Grants Numbers MOST 104-2112-M-006-016-MY3 and
105-2112-M-005-008-MY3, respectively. F.N. was also partially
supported by the RIKEN iTHES Project, the MURI Center for Dynamic
Magneto-Optics via the AFOSR award number FA9550-14-1-0040, the
IMPACT program of JST, and a Grant-in-Aid for Scientific Research
(A).
\end{acknowledgments}
%------------------------------------------------------------------
%\bibliography{bib_sts}


\appendix

\section{Spatio-Temporal Steering for Testing Nonclassical Correlations in Quantum Networks: Supplementary Material}
We here present the model Hamiltonian for the Fenna-Matthews-Olson
(FMO) protein complex, and the corresponding Haken-Strobl equation
which describes the open-system excitation transfer in the FMO
complex. The decay of the spatio-temporal steering robustness,
studied in Example 2 of the main article, was calculated for a
numerical solution of this equation.

The model Hamiltonian of the single FMO monomer containing $N$
sites can be written as (see, e.g. Ref.~\cite{Chen13} and
references therein):
\begin{equation}
H=\sum_{n=1}^{N}\frac{\epsilon _{n}}{2}\sigma_z^{(n)}+\sum_{n<n^{\prime
}}J_{n,n^{\prime }}(\sigma_+^{(n)}\sigma_-^{(n^{\prime })}+\sigma_-^{(n)}\sigma_+^{(n^{\prime })})
\end{equation}%
where the state Pauli operators represent an electronic excitation
at site $n$, ($n\in $ 1,...,7), such that
$\sigma_z^{(n)}=|e^{(n)}\rangle\langle e^{(n)}
|-|g^{(n)}\rangle\langle g^{(n)}|$, $\epsilon _{n}$ is the site
energy of chromophore $n$, and $ J_{n,n^{\prime }}$ is the
excitonic coupling between the $n$th and $n^{\prime }$th sites. In
the literature, because of the rapid recombination of multiple
excitations in such a complex, it is common to simplify
drastically this model by assuming that the whole complex only
contains a single excitation. In that case the $2^7$ dimensional
Hilbert space is reduced to a $7$ dimensional Hilbert space. Here,
while we also assume only a single-excitation, we keep the full
$2^7$ dimensional Hilbert space to enable us to consider
measurements in a basis which represent superpositions of
excitations at various sites. (Note that for simplicity, we omit
the recently discovered eighth site~\cite{olbrich}).

 In the regime that the excitonic
coupling $J_{n,n^{\prime }}$ is large compared with the
reorganization energy, the electron-nuclear coupling can be
treated perturbatively~\cite{aki}, and the open-system dynamics of
the system can be described by the Haken-Strobl master-type
equation~\cite{Haken,MohseniBook},
\begin{equation}
\dot{\rho}(t)=-\frac{i}{\hbar }[H,\rho ]+L[\rho ],
\end{equation}%
where $\rho $ is the system density matrix, and $L[\rho ]$ denotes
the Lindblad operators
\begin{equation}
L[\rho ]=L_{\text{sink}}[\rho ]+L_{\text{deph}}[\rho ],
\end{equation}%
where the Lindblad superoperator $L_{\text{sink}}$ describes the
irreversible excitation transfer from site-3 to the reaction
center:
\begin{equation}
L_{\text{sink}}[\rho ]=\Gamma \lbrack 2s\rho s^{\dagger }-s^{\dagger }s\rho
-\rho s^{\dagger }s],
\end{equation}%
where $s=\sigma_+^{(R)}\sigma_-^{(3)}$, with $\sigma_+^{(R)} $
representing the creation of an excitation in the reaction center,
and $\Gamma $ denotes the transfer rate. The other Lindblad
superoperator, $L_{\text{deph}}$, describes the
temperature-dependent dephasing with the rate $\gamma
_{\text{dp}}$:
\begin{equation}
L_{\text{deph}}[\rho ]=\gamma _{\text{dp}}\sum_{n}[2A_{n}\rho A_{n}^{\dagger
}-A_{n}A_{n}^{\dagger }\rho -\rho A_{n}A_{n}^{\dagger }],
\end{equation}%
where $A_{n}=\sigma_z^{(n)}$. This dephasing Lindblad operator
leads to the exponential decay of the coherences between different
sites in the system density matrix. The pure-dephasing rate
$\gamma_{\textrm{dp}}$ can be estimated by applying the standard
Born-Markov system-reservoir model~\cite{open, seth}.  We assume
an Ohmic spectral density, which, combined with the Born-Markov
approximations, leads to a dephasing rate directly proportional to
the temperature \cite{seth}.  While more complex treatments are
necessary to fully describe the true dynamics of the FMO complex,
here we restrict ourselves to this weak-coupling Lindblad form for
numerical efficiency and easier interpretation of results.
Note that there exists a factor $1/8$ between the dephasing rate
$\gamma_{\text{dp}}$ here and that in the orthodox seven-site FMO
model.

In the FMO monomer, the excitation transferring from site-3 to the
reaction center takes place on a time scale of $\sim 1$ ps, and
the dephasing occurs on a time scale of $\sim 100$ fs \cite{seth}.
These two time scales are both much faster than that of the
excitonic fluorescence relaxation ($\sim 1$ ns), which is, thus,
omitted here for simplicity. Here we present the values used for
the system Hamiltonian in calculating the excitation
transfer~\cite{Renger}:
\begin{widetext}
\begin{eqnarray} \label{Hamiltonian2}
H'= \left(%
\begin{array}{ccccccc}
215 & -104.1 & 5.1 & -4.3 & 4.7 & -15.1 & -7.8 \\
-104.1 & 220 & 32.6 & 7.1 & 5.4 & 8.3 & 0.8 \\
5.1 & 32.6 & 0 & -46.8 & 1.0 & -8.1 & 5.1 \\
-4.3 & 7.1 & -46.8 & 125 & -70.7 & -14.7 & -61.5 \\
4.7 & 5.4 & 1.0 & -70.7 & 450 & 89.7 & -2.5 \\
-15.1 & 8.3 & -8.1 & -14.7 & 89.7 & 330 & 32.7 \\
-7.8 & 0.8 & 5.1 & -61.5 & -2.5 & 32.7 & 280\notag\\
\end{array}
\right)
\end{eqnarray}
\end{widetext}
Here the diagonal elements correspond to $\epsilon _{n}$, and the off-diagonals to $J_{n,n^{\prime }}$. We omit the large ground-state off-set, as it does not influence the results.  This FMO dynamics description is based on our former work~\cite{Chen13}.

\end{document}